\documentclass[twocolumn,prl]{revtex4-1}
\usepackage{graphicx}% Include figure files
\usepackage{float}

\begin{document}

\title{Field-Induced Magnetic Monopole Plasma in Artificial Spin Ice}

\author{M. Goryca$^1$, X. Zhang$^2$, J. Li$^1$, A. L. Balk$^1$, J. D. Watts$^{3,4}$, C. Leighton$^3$, C. Nisoli$^5$, P. Schiffer$^{2,6}$, S. A. Crooker$^1$}

\affiliation{$^{1}$National High Magnetic Field Lab, Los Alamos National Laboratory, Los Alamos, NM 87545, USA}
\affiliation{$^{2}$Department of Applied Physics, Yale University, New Haven, CT 06520, USA}
\affiliation{$^{3}$Department of Chemical Engineering and Materials Science, University of Minnesota, Minneapolis, MN 55455, USA}
\affiliation{$^{4}$School of Physics and Astronomy, University of Minnesota, Minneapolis, MN 55455, USA}
\affiliation{$^{5}$Theoretical Division, Los Alamos National Laboratory, Los Alamos, NM 87545, USA}
\affiliation{$^{6}$Department of Physics, Yale University, New Haven, CT 06520, USA}

\begin{abstract}
Artificial spin ices (ASIs) are interacting arrays of lithographically-defined nanomagnets in which novel frustrated magnetic phases can be intentionally designed. A key emergent description of fundamental excitations in ASIs is that of \textit{magnetic monopoles} -- mobile quasiparticles that carry an effective magnetic charge. Here we demonstrate that the archetypal square ASI lattice can host, in specific regions of its magnetic phase diagram, high-density plasma-like regimes of mobile magnetic monopoles. By passively ``listening'' to spontaneous monopole noise in thermal equilibrium, we reveal their intrinsic dynamics and show that monopole kinetics are minimally correlated (that is, most diffusive) in the plasma phase. These results open the door to on-demand monopole regimes having field-tunable densities and dynamic properties, thereby providing a new paradigm for probing the physics of effective magnetic charges in synthetic matter.
\end{abstract}

\maketitle

Owing to their user-defined geometries of interacting magnetic elements, artificial spin ices (ASIs) provide a highly flexible and powerful platform with which to investigate the rich physics of frustrated spin systems~\cite{Skjaervo2020, Rougemaille2019, Gilbert2016_PhysToday}. Initially conceived~\cite{Wang2006} as two-dimensional analogs of ``natural'' pyrochlore spin ices~\cite{Harris1997, Bramwell2020} such as Ho$_2$Ti$_2$O$_7$, investigations of ASIs now extend well beyond these original goals and enable detailed studies of a vast selection of possible interacting lattice arrangements, including exotic magnetic topologies not found in nature~\cite{Nisoli2017, Farhan2017, Drisko2017, Wang2016}. Together with natural spin ice materials, one of their most exciting properties is that the fundamental excitations in many ASIs have a natural emergent description in terms of effective magnetic monopoles~\cite{Castelnovo2008, Bramwell2009} -- that is, mobile quasiparticles that possess the equivalent of a net magnetic charge.  These charge excitations can interact with each other and with applied magnetic fields via the magnetic analog of the ubiquitous electronic Coulomb interaction, representing the emergence of a range of novel phenomena~\cite{Skjaervo2020, Rougemaille2019}, including the possibility of ``magnetricity''~\cite{Bramwell2009}. 

While the presence of monopoles in ASI has been observed in pioneering imaging measurements~\cite{Ladak2010, Mengotti2010}, dynamical studies of monopole kinetics, and the ability to tune continuously through monopole-rich phases, remain at an early stage. Because their underlying magnetic interactions can be engineered to manifest near room temperature, ASIs are especially well-suited to studies of monopole dynamics and other collective modes.  In this work, we use a high-bandwidth magneto-optical noise spectrometer to passively detect spontaneous magnetization fluctuations in thermally active square ASI. Because fluctuations of the constituent magnetic elements in ASIs are inextricably linked to the kinetics of monopoles, the system's broadband magnetization noise spectrum naturally encodes the intrinsic timescales and dynamic correlations of the underlying monopole excitations. The noise reveals specific regions in the field-dependent phase diagram where the density of mobile monopoles increases well over an order of magnitude compared with neighboring phases, a consequence of the field-tunable tension on the Dirac strings connecting mobile monopoles. Moreover, detailed noise spectra demonstrate that monopole kinetics are minimally correlated (\textit{i.e.}, most diffusive) in this plasma-like regime. Discovery of on-demand monopole phases with tunable kinetic properties opens the door to new probes of magnetic charge dynamics and provides a new paradigm for studies of magnetricity in artificial magnetic materials.

We consider the prototypical square ASI lattice~\cite{Wang2006}, shown in Fig. 1(a).  Each ferromagnetic nano-island behaves as a single Ising-like macrospin with net magnetization parallel or antiparallel to its long axis due to shape anisotropy. Crucially, the islands are made sufficiently thin so that they are superparamagnetic and thermally active at room temperature~\cite{Morgan2011,Budrikis2012,Kapaklis2014,Farhan2013,Anghinolfi2015}, \textit{i.e.}, in the absence of strong biasing magnetic fields, each island's magnetization can fluctuate spontaneously. This ensures that the lattice can efficiently sample the vast manifold of possible moment configurations, and remain near its magnetic ground state in thermal equilibrium.

\begin{figure*} 
\center
\includegraphics[width=.98\textwidth]{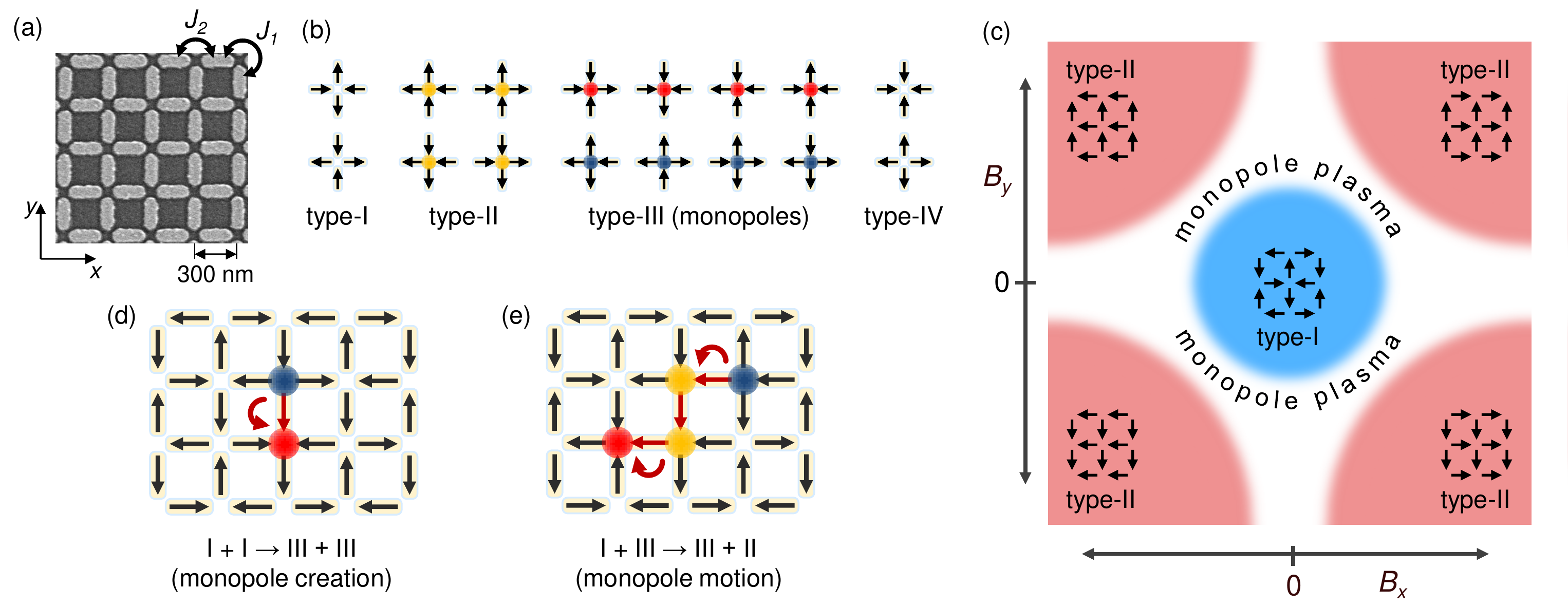}
\caption{Field-dependent phase diagram of thermally-active square artificial spin ice (ASI). (a) An SEM image of the sample.  Each Ni$_{0.8}$Fe$_{0.2}$ island has lateral dimensions 220~nm $\times$ 80~nm and thickness 3.5~nm, and behaves as a single superparamagnetic Ising moment that, in the absence of any biasing field, exhibits rapid thermodynamic fluctuations near room temperature. (b) The four vertex types in archetypal square ASI, in order of increasing energy at zero applied magnetic field.  Type-I vertices have lowest energy because nearest-neighbor dipolar coupling $J_1$ exceeds the next-nearest-neighbor coupling $J_2$. Type-I and -II vertices have 2-in/2-out configurations and therefore obey ice rules (but only type-II have a net polarization), while type-III vertices have 3-in/1-out or 3-out/1-in configurations and therefore have a monopole-like effective magnetic ``charge''. (Type-IV vertices also have magnetic charge but are energetically very unfavorable and occur only rarely.) (c) Notional schematic of the anticipated field-dependent phase diagram of square ASI, showing the two well-defined ground states of the system: full tiling with type-I vertices at small applied fields $B_{x,y} \approx 0$, and polarized type-II vertex tiling when $|B_x|$ and $|B_y|$ are both large. Near the boundaries, the equilibrium dynamics are determined by the thermal creation, annihilation, and motion of type-III monopole vertices, which generate magnetization noise. (d) A thermal fluctuation in the type-I phase creates a pair of type-III vertices. (e) Subsequent fluctuations can cause the monopoles to diffuse, leaving behind a Dirac string of type-II (yellow) vertices. Red arrows show islands that have flipped; blue and red dots indicate the mobile monopole-like vertices.}
\label{fig1}
\end{figure*}

Our study focuses on a previously-unexplored characteristic of thermal square ASI; namely, that its field-dependent magnetic phase diagram and ground-state moment configuration \textit{must} include regions where monopole-like excitations play a dominant and active role. This can be understood by considering the relative energies of the four possible vertex types (I-IV), shown and described in Fig. 1(b) in order of increasing energy at zero applied magnetic field.  At zero field, the ground state of square ASI is an ordered antiferromagnetic tiling of type-I vertices~\cite{Morgan2011,Budrikis2012,Zhang2013}, which obey the ``2-in/2-out'' ice rule and do not possess any net polarization. However, for sufficiently large in-plane magnetic fields applied at angles near a lattice diagonal ($\pm$45$^\circ$), polarized type-II vertices must eventually become energetically favored; these also obey ice rules, but possess a net polarization. The field-dependent phase diagram of thermal square ASI should then qualitatively resemble the schematic drawn in Fig. 1(c). 

The crossover between type-I and type-II magnetic order (or between type-II orderings with different polarization) obviously requires the reversal of individual islands.  Crucially, as depicted in Fig. 1(d), flipping any island within either a type-I or type-II ordered lattice unavoidably creates a pair of higher-energy type-III vertices, which have 3-in/1-out or 3-out/1-in moment configuration and therefore possess an effective magnetic ``charge'' that can be regarded as a magnetic monopole-like quasiparticle excitation~\cite{Skjaervo2020,Rougemaille2019, Gilbert2016_PhysToday}. Subsequent flips of other islands can create additional monopole pairs, annihilate adjacent monopole pairs, or cause an existing monopole to move through the ASI lattice (Fig. 1(e)). Within a type-I (type-II) ordered region, the creation and subsequent separation of a monopole pair along a staggered diagonal direction creates a string of type-II (type-I) vertices~\cite{Farhan2013}. As discussed in detail below, near the type-I/type-II boundaries these monopole excitations, once thermally created, can diffuse \textit{freely} along certain directions with no cost in energy. Thermal square ASI may therefore be expected to host field-tunable regimes of mobile magnetic monopoles (see Fig. 1(e)). 

To search for dynamic monopole regimes in square ASI, and to quantify their timescales and correlations --all under conditions of strict thermal equilibrium-- we developed a broadband magnetization noise spectrometer to measure the frequency spectrum of the system's intrinsic magnetization fluctuations (see Fig. 2(a), and Appendix). Samples were mounted in the $x-y$ plane, with horizontal and vertical islands oriented along $\hat{x}$ and $\hat{y}$. A weak probe laser was linearly polarized and focused on the ASI. Due to the longitudinal magneto-optical Kerr effect (MOKE), magnetization fluctuations in the $\hat{x}$ direction, $\delta m_x(t)$, imparted Kerr rotation fluctuations $\delta \theta_K(t)$ on the polarization of the reflected laser, which were detected with balanced photodiodes. This ``magnetization noise'' was digitized and its frequency-dependent power spectrum $P(\omega)$ was computed and signal-averaged in real time. Figure 2(b) shows a characteristic spectrum measured out to 1~MHz, spanning over 5 orders of magnitude in both frequency and detected noise.  

\begin{figure} 
\center
\includegraphics[width=.5\textwidth]{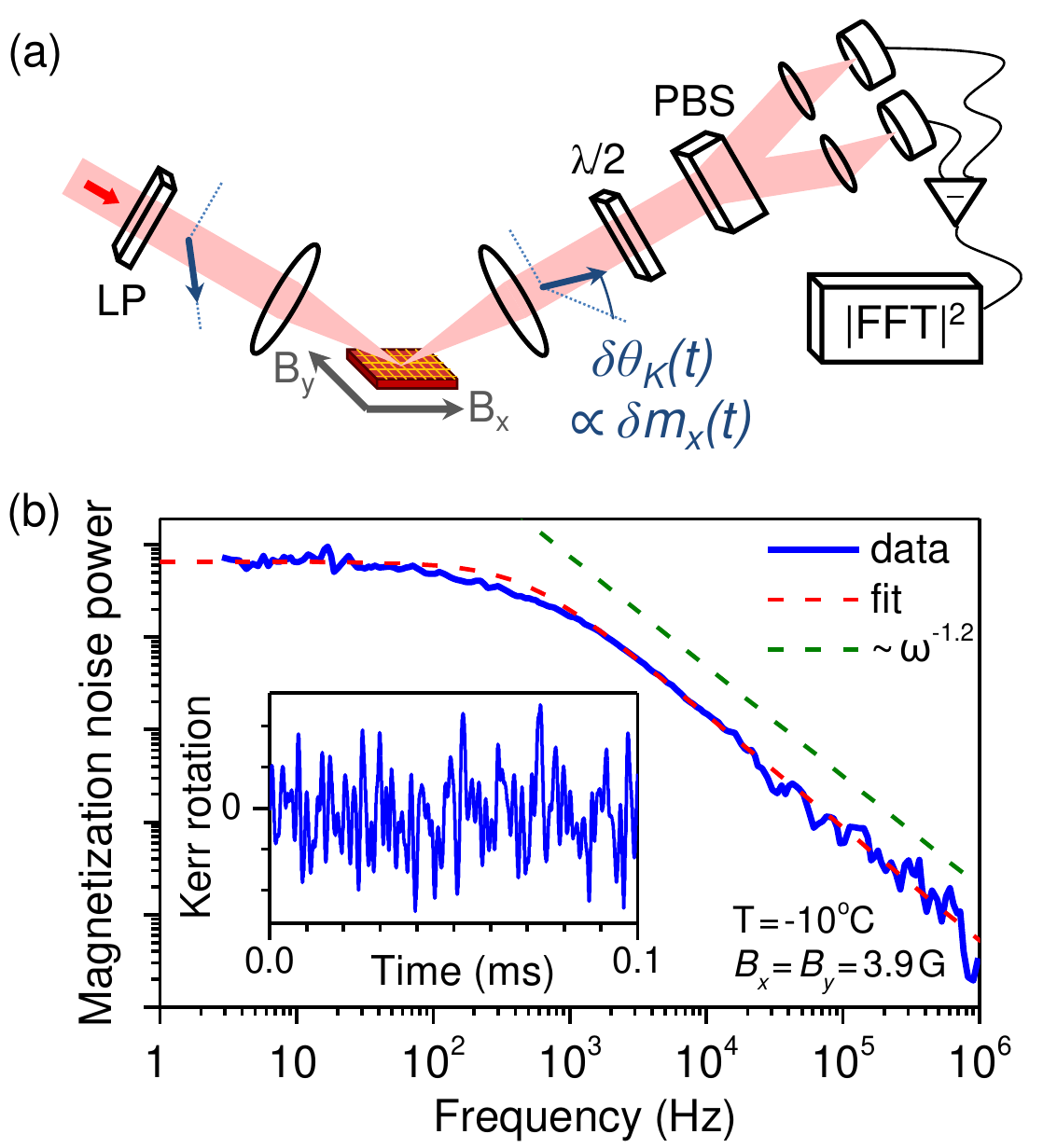}
\caption{Optical detection of broadband magnetization noise in ASI. (a) Experimental schematic: A weak probe laser is linearly polarized (LP) and focused on the ASI.  Magnetization fluctuations of the horizontal islands, $\delta m_x (t)$, impart Kerr rotation fluctuations $\delta\theta_{K}(t)$ on the reflected laser which are detected with a polarization beamsplitter (PBS) and balanced photodiodes. The frequency spectrum of the magnetization noise power, $P(\omega)$ -- equivalent to the Fourier transform of the temporal magnetization correlation function $\langle \delta m_x(0)~\delta m_x(t) \rangle$ -- is computed and averaged in real time. (b) An example of characteristic magnetization noise from square thermal ASI in the time domain (inset), and in the corresponding frequency domain spanning nearly six orders of magnitude from 1~Hz to 1~MHz. The dashed red line shows a fit to the model described in the main text, and the dashed green line shows that the noise decays as a power law at high frequencies ($P \sim \omega^{-1.2}$), indicating correlated (sub-diffusive) kinetics.}
\label{fig2}
\end{figure}

We first analyze the noise from a control lattice that contains only horizontal islands oriented along $\hat{x}$. Figure 3(a) shows a map of the measured \textit{total} (frequency-integrated) noise power versus applied in-plane magnetic fields $B_x$ and $B_y$. As expected, significant noise was only observed when $B_x \approx 0$, where the thermally-active islands were unbiased and fluctuated freely. For larger $|B_x|$, the islands were polarized and effectively frozen, suppressing fluctuations. Note that $B_y$ has no effect because this control sample lacks vertical islands, and because the Ising-like magnetization of the horizontal islands is not influenced by $B_y$ in this small range ($\pm$7~G). 

In marked contrast, Fig. 3(b) shows the noise map from square ASI, where strong dipolar interactions between nearest-neighbor (adjacent vertical and horizontal) islands lead to the stable type-I magnetic ordering sketched in the phase diagram of Fig. 1(c).  Indeed, the noise map exhibits a large dark central region when $B_{x,y} \approx 0$, indicating suppressed fluctuations, as expected for stable type-I order. Towards the four corners of the map, where both $|B_x|$ and $|B_y|$ are large, the magnetization noise again vanishes, indicating field-stabilized type-II tiling. Strikingly, however, there is a bright diamond-shaped boundary region, indicating a high level of spontaneous noise between the type-I and type-II magnetic orderings.  

\begin{figure*} 
\center
\includegraphics[width=.76\textwidth]{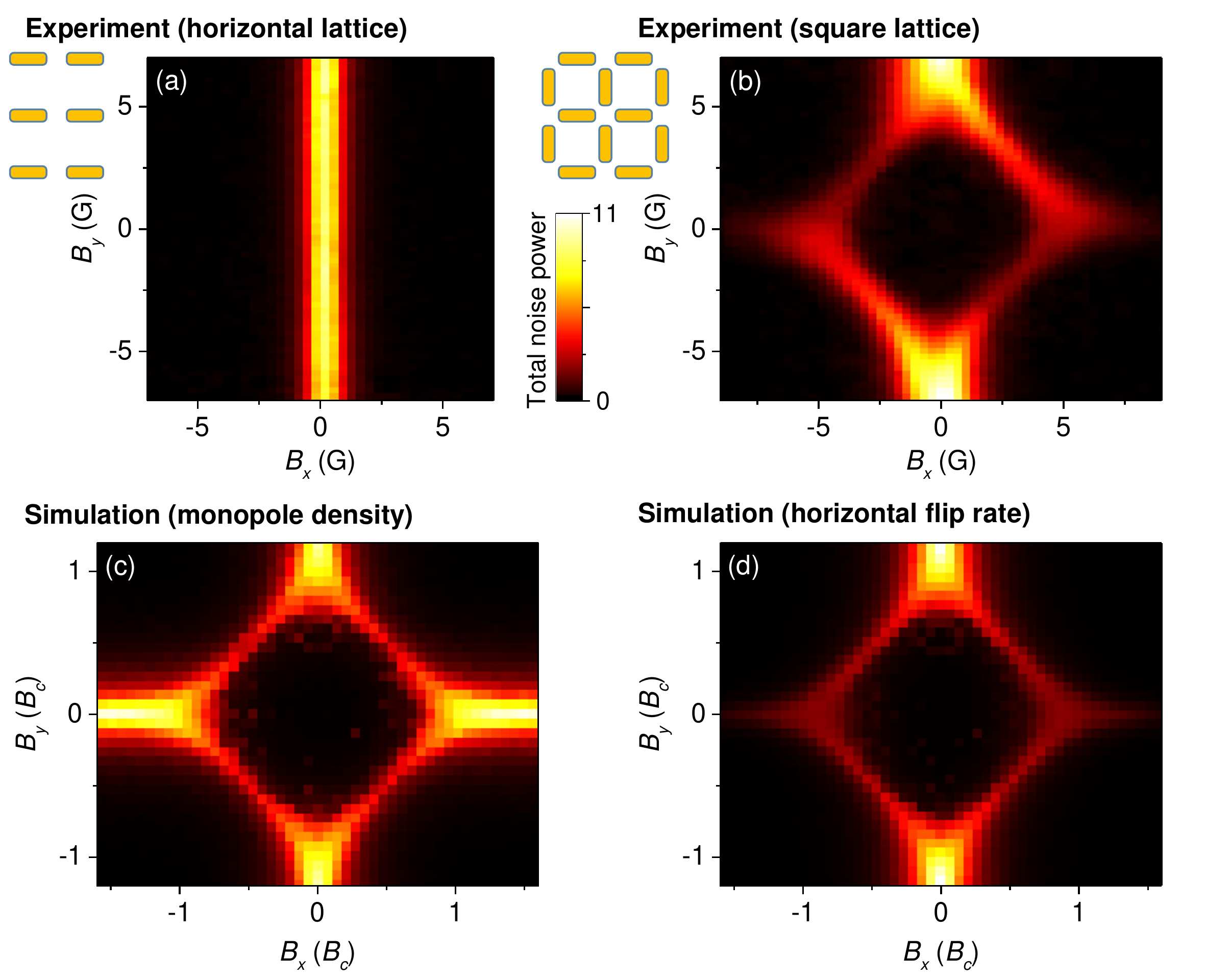}
\caption{Noise maps reveal the magnetic phase diagram of thermally-active square ASI. (a) A map of the total magnetization noise power ($\int P(\omega) d\omega$, integrated from 250~Hz to 250~kHz) from a control sample that contains only horizontal islands, versus applied in-plane magnetic fields $B_x$ and $B_y$. $T=-10.0^{\circ}$C. For $B_x \approx 0$, the islands fluctuated spontaneously, but fluctuations were suppressed at larger $\pm B_x$ where all the islands were polarized by the field. (b) A corresponding noise map from square ASI. The diamond-shaped boundary, where type-I and type-II vertices are degenerate in energy, is revealed by thermodynamic noise arising from the dynamics of type-III (monopole-like) vertices. The map lacks the four-fold symmetry of the square lattice because it measures only fluctuations along $\hat{x}$ but not $\hat{y}$. (c) A corresponding map of the calculated density of monopole vertices from Monte Carlo spin dynamics simulations. Along 
the diagonal boundary where the density is large, the majority of the monopoles are free (see Fig. 4). (d) A map of the calculated rate of horizontal magnetization flips. Measured and simulated noise maps at $T=6^{\circ}$C and $23^{\circ}$C, and measured noise maps for square ASI with a different lattice constant, are shown in Supplemental Material Figs. S2 and S3.}
\label{fig3}
\end{figure*}

The boundary region occurs when $|B_x| + |B_y|$ equals the crossover field $B_c$ where type-I and type-II vertices become exactly degenerate in energy (see  Supplemental Material Fig. S1 for additional details). As discussed above, spontaneous reversal of an island within a type-I or type-II ordered region creates a pair of type-III monopole vertices (Fig. 1(d)). The key point is that at $B_c$, these thermally-generated monopoles can then separate and diffuse \textit{freely}  along the staggered diagonal path that is most closely aligned with the applied field, as illustrated in Fig. 1(e). This motion, denoted by the process $\mathrm{III} + \mathrm{I} \leftrightarrow \mathrm{II} +\mathrm{III}$, has no collective energy cost along the boundary where type-I and -II vertex energies are degenerate. No net energy is required to lengthen or shorten the Dirac string of flipped vertices connecting the two monopoles -- \textit{i.e.}, the string tension vanishes~\cite{Farhan2013}. Most importantly, the freedom to separate leads to a substantial increase in the effective monopole lifetime and therefore the equilibrium density of mobile monopole vertices. Being topological quasiparticles, their density is limited only by the rate at which they annihilate by diffusing to an edge of the lattice or by encountering monopoles of opposite charge.  This special regime stands in marked contrast to the case when $|B_x| + |B_y| <B_c$ (or $>B_c$), where monopole motion that creates additional type-II (or type-I) vertices is energetically unfavored and therefore suppressed.  In this case the Dirac string has a non-zero tension that favors recombination of monopole pairs shortly following their creation, and, as demonstrated below, is accompanied by increasingly anomalous monopole diffusion and correlated dynamics. 

Because the stochastic creation, annihilation, and motion of monopoles is intimately linked to reversal of individual islands, the boundary region is therefore clearly revealed by magnetization noise.  Equivalently, all noise in square ASI is necessarily due to monopole kinetics. The diamond-shaped boundary therefore signals a field-tunable regime of dynamic magnetic monopoles. We note that the noise map also shows bright vertical stripes when $B_x \approx 0$ and $|B_y|$ is large, indicating strong fluctuations at the boundary between different type-II orientations. This arises from an effective dimensional reduction, where all the vertical islands are polarized by $B_y$ and therefore --considering nearest-neighbor coupling $J_1$ only-- there is no energy difference whether a horizontal island is magnetized along $\pm \hat{x}$. Thermal fluctuations $\delta m_x(t)$ therefore occur, similar to the case of the control lattice at $B_x=0$. (Weaker next-nearest-neighbor coupling $J_2$ is insufficient, compared to $kT$, to stabilize any order).   

To better understand the noise data and provide additional insight into the underlying mechanisms involved, we performed Monte Carlo simulations of square ASI (see Appendix). Figure 3(c) shows the calculated density of type-III monopole vertices, while the computed rate of horizontal spin flips is shown in Fig. 3(d). Regions of high monopole density correspond to regions of maximum flip rate, which in turn agrees very well with the measured noise map, thereby validating the connection between the measured noise power and the density of mobile monopoles. 

\begin{figure} 
\center
\includegraphics[width=0.5\textwidth]{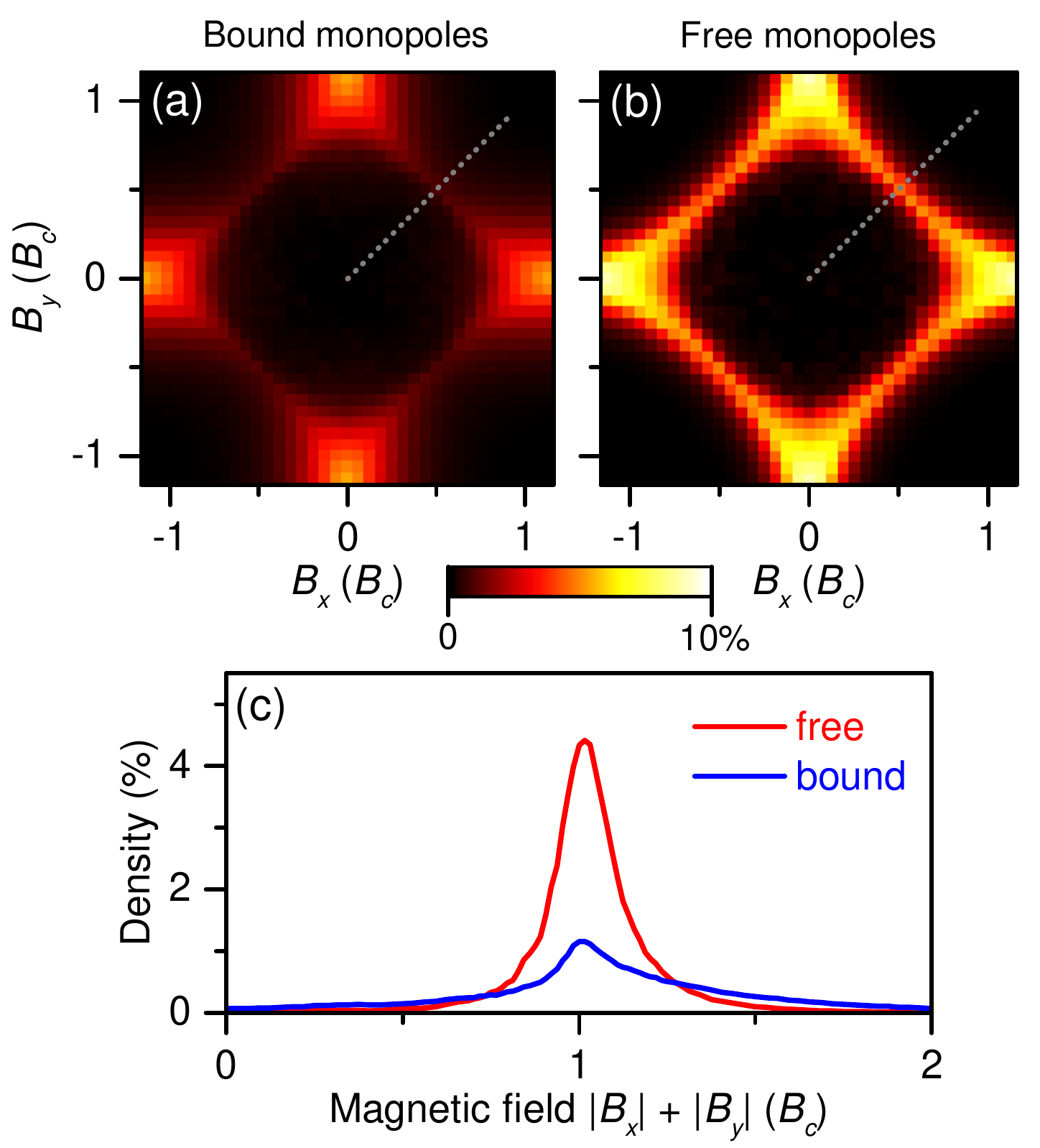}
\caption{Calculated densities of ``bound'' and ``free'' monopoles \emph{vs.} in-plane magnetic field, from Monte Carlo simulations. (a) Map of the density of ``bound'' monopoles (defined as type-III vertices that have a neighboring type-III vertex of opposite magnetic charge) \emph{vs.} magnetic field, simulated at $T=1.2$ $J_2/k$. (b) Corresponding map of ``free'' (not bound) monopole density. (c) Bound and free monopole densities \emph{vs.} magnetic field applied along a lattice diagonal direction (\emph{i.e.}, along the dotted lines in panels (a) and (b)). In the monopole plasma regime (\emph{i.e.}, near the crossover field $B_c$ where type-I and type-II vertices are energetically degenerate), the total monopole density increases mainly due to the huge increase in the number of free monopoles, which in turn is due to the absence of Dirac string tension at $B_c$ and consequent free monopole diffusion.}
\label{fig4}
\end{figure}

Returning to the phase diagram sketched in Fig. 1(c), both experiments and simulations confirm that the type-I and type-II ordered regions in square ASI are separated by a plasma-like regime of thermally-active monopoles. Simulations indicate that not only does the monopole density increase by over two orders of magnitude at these boundaries (Fig. 3(c)), but also that the majority of these monopoles are freely diffusing (Fig. 4). Crucially, we note that both the monopole density and (as shown below) their dynamic correlations in this regime are continuously tunable by applied magnetic fields above and below $B_c$, which is distinct from the monopole phase recently achieved in ``degenerate square ASI'' that uses height-offset vertical and horizontal islands~\cite{Perrin2016,Farhan2019}.  In other words, by tuning to $B_c$ in conventional square ASI, we realize a regime where correlations associated with the energy costs of changing type-I and type-II vertex populations are minimized.  Of course, tuning $|B|$ away from $B_c$ lifts the energy degeneracy of type-I and type-II vertices, whereupon the Dirac strings acquire tension and the monopoles will be affected by the magnetostatic potential associated with $B$, and can be expected to exhibit different kinetics. 

Noise measurements provide an effective tool to directly probe kinetic correlations of monopoles, via the detailed shape of the noise spectrum over a very broad frequency range (typically from 1~Hz to 1~MHz). This range directly accesses the relevant intrinsic timescales of fluctuations in our thermally-active ASIs, and naturally complements powerful imaging techniques such as PEEM or MFM that provide excellent spatial information but are typically limited to much slower ($\sim$0.1-0.001~Hz) timescales and are less amenable to measurements in applied magnetic fields. We note that an analogous approach based on SQUID magnetometry was recently applied to Dy$_2$Ti$_2$O$_7$ crystals by Dusad \textit{et al.}~\cite{Dusad2019}, yielding important insights into monopoles in natural pyrochlore spin ices at cryogenic temperatures.

\begin{figure*}[t]
\center
\includegraphics[width=.8\textwidth]{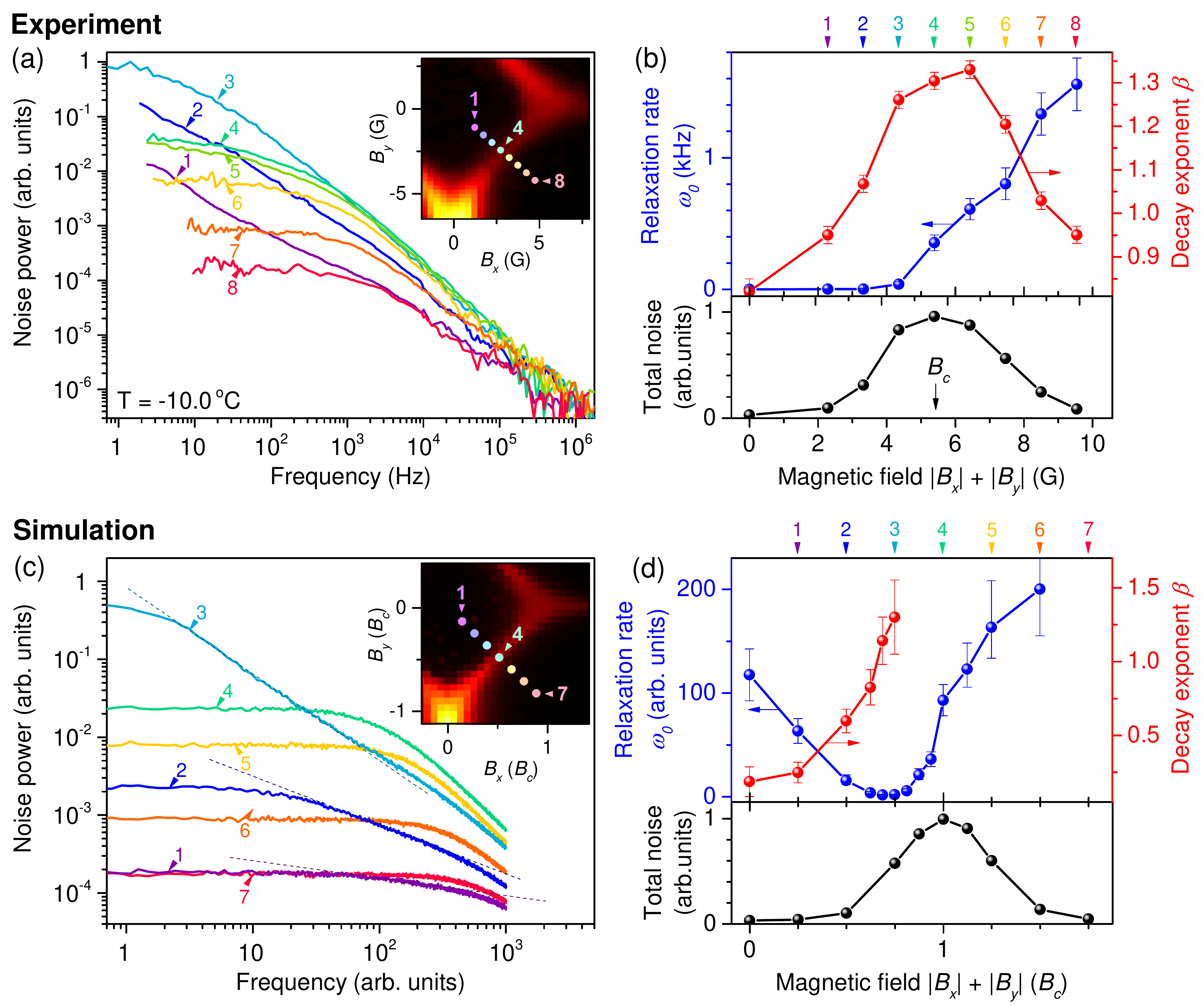}
\caption{Magnetization noise spectra through the monopole plasma regime. (a) Measured noise spectra along the indicated path (keeping $|B_x|=|B_y|$), which traverses between type-I and type-II order. $T=-10.0^{\circ}$C. (b) Extracted  relaxation rates $\omega_0$ (blue points), decay exponents $\beta$ (red points), and total noise power (black points). $\beta$ exhibits a maximum and is closest to a value of 2 when $B = B_c$, indicating that the monopoles exhibit their most-uncorrelated (most diffusive) dynamics in the plasma-like regime where their density is largest. (c), (d) Corresponding computed noise spectra from Monte Carlo simulations and extracted parameters $\omega_0$, $\beta$, and total noise. The total integrated noise is peaked when type-I and type-II vertex energies are degenerate, \textit{i.e.}, in the monopole plasma regime that occurs at the crossover field $B_c$. By contrast, simulations show that $\omega_0$ exhibits a minimum when the ASI orders antiferromagnetically (spontaneous type-I ordering), which occurs at an applied field that is slightly less than $B_c$ when $T>0$ (see also Supplemental Material Fig. S4).}
\label{Crossover}
\end{figure*}

A typical noise spectrum, $P(\omega)$, from square ASI is shown in Fig. 2(b). Here, $P(\omega) = \langle a(\omega) a^*(\omega) \rangle$, where $a(\omega)$ is the Fourier transform of the noise signal $\delta m(t)$ and the brackets indicate an average over repeated measurements. Equivalently, $P(\omega)$ is the Fourier transform of the system's temporal correlation function $\langle \delta m(0) \delta m(t) \rangle$. Empirically, we find that all the measured noise spectra can be fit reasonably well by the functional form
\begin{equation}
P(\omega) \propto  \frac{1}{(\omega^2+\omega_0^2)^{\beta/2}},
\label{fit}
\end{equation}
where $\omega_0$ is a characteristic relaxation rate, and $\beta$ is a power-law decay exponent where $P(\omega) \propto \omega^{-\beta}$ at high frequencies. Importantly, $\beta$ defines the ``color'' of the noise and is an indicator of correlated dynamics. In general, if processes responsible for magnetization dynamics are uncorrelated in time, then $\langle \delta m(0) \delta m(t) \rangle$ decays exponentially ($\propto e^{-t/\tau_0}$) with a characteristic relaxation time $\tau_0 = 1/ \omega_0$. Per the fluctuation-dissipation theorem, the corresponding noise spectrum then exhibits a Lorentzian lineshape ($\beta = 2$) and is said to be Brownian. Noise exhibiting $\beta=2$ therefore implies simple diffusive Brownian kinetics, such as from a trivial random walk with independent increments, or from monopole creation and annihilation models described, \textit{e.g.}, by Ryzhkin~\cite{Ryzhkin2005} and Klyuev~\cite{Klyuev2019}.  In contrast, noise exhibiting $\beta < 2$ indicates kinetics that cannot be described by a single exponential time scale~\cite{Bovo2013}. Although it is always possible to simulate any $\beta$ by summing a suitable distribution of Lorentzians, for the case of ASI the observation of $\beta<2$ can be considered in terms of anomalous monopole diffusion arising from correlated kinetics.  In particular, when temporal fluctuations are no longer independent but instead exhibit negative correlations (\textit{e.g.}, if prior fluctuations increased the magnetization, then the next fluctuation is more likely to decrease it), then monopoles will exhibit sub-diffusive behavior. Such processes are related to ``fractional'' Brownian motion~\cite{Mandelbrot1968} and can be said to retain a memory~\cite{Klyuev2019}. Noise following Eq. (1) can derive from kinetics with algebraically-decaying correlation functions of the form $\langle \delta m(0) \delta m(t) \rangle  \propto t^{-\alpha} e^{-t/\tau_0}$, where $\alpha = 1- \frac{\beta}{2}$. Thus, broadband noise spectroscopy can reveal both the kinetic correlations and the characteristic relaxation times as square ASI is tuned through the monopole plasma regime. 

Figure 5(a) shows the dramatic evolution of $P(\omega)$ as square ASI is tuned across the boundary between type-I and type-II order, keeping $|B_x|$ = $|B_y|$.  Figure 5(b) shows the measured values of $\omega_0$, $\beta$, and the total noise power. Interestingly, although $\beta < 2$ everywhere (indicating some degree of correlated dynamics), $\beta$ exhibits a clear maximum and is closest to 2 when $B=B_c$, where the total noise power and the monopole density are largest. This indicates that memory effects are weakest in the plasma-like regime, and monopole motion along the lattice diagonal ($\mathrm{III} + \mathrm{I} \leftrightarrow \mathrm{II} +\mathrm{III}$; see Fig. 1(e)) most closely approximates ordinary diffusion. However, $\beta$ decreases when $|B| \neq B_c$, indicating that dynamics in square ASI become increasingly sub-diffusive away from the monopole plasma regime. Given that noise in pyrochlore Dy$_2$Ti$_2$O$_7$ also evinced similarly sub-diffusive kinetics~\cite{Dusad2019}, correlated monopole dynamics in thermal equilibrium may therefore be a universal feature shared by both natural (3D) and artificial (2D) spin ices.  

Monte Carlo simulations capture the overall shape of the noise spectra and many of the observed trends (Figs. 5(c),(d)), showing the total noise is maximized at the monopole plasma regime at $B_c$. The simulated noise also exhibits a non-trivial power-law decay exponent $\beta$ in the type-I phase, that grows steeper (see dashed lines) as $B$ increases toward $B_c$, in agreement with the data.  The simulations do not permit accurate extraction of $\beta$ at larger $B$ (\textit{i.e.}, into the type-II phase), because of the proximity of $\omega_0$ to the highest (Nyquist) frequency of the simulation, where aliasing artifacts and trivial decays due to numerics are also present. 

The measured spectra (Figs. 5(a),(b)) also show that the relaxation rate $\omega_0$ falls by orders of magnitude when tuning from type-II into the type-I phase by decreasing $|B|$. Such behavior is only partially expected:  While a slowing-down of kinetics upon approaching the antiferromagnetic phase transition is expected from theory~\cite{Hohenberg1977,Chen2007}, relaxation rates are expected to increase once again after entering the antiferromagnetically-ordered  phase -- as captured by the simulations (Figs. 5(c),(d)) but in marked contrast to the measurements where $\omega_0$ remains small as $B \rightarrow 0$.  This discrepancy points to a key difference between real ASIs, which are composed of mesoscopic superparamagnets, and models of binary spin systems. Kinetics in real ASIs are a non-trivial convolution of many-body time scales (which are simulated), and local timescales associated with magnetic anisotropy and dynamics within the islands themselves (which typically are not simulated). 

The calculated minimum in $\omega_0$ highlights a further important distinction: the antiferromagnetic phase transition in square ASI~\cite{Sendetskyi2019} occurs at $B<B_c$ when $T>0$ (as mandated by the system's temperature-dependent free energy), whereas the monopole plasma always occurs at $B_c$ --independent of temperature-- because it is determined solely by the energy degeneracy of type-I and type-II vertices.  This distinction is apparent in the separation between $B_c$ and the minimum in $\omega_0$ in Fig. 5(d), and is further elucidated at other temperatures by calculations of the specific heat and order parameter shown in Supplemental Material Fig. S4.

In summary, broadband noise spectroscopy introduces a new paradigm to ASI studies, by providing a probe that is explicitly sensitive to dynamic timescales and correlations over many orders of magnitude in frequency.  These results open the door to direct exploration of field/temperature phase diagrams and their intrinsic equilibrium dynamics, and the discovery of a field-tunable monopole plasma regime in archetypal square ASI demonstrates the power of such investigations. The ability to create monopole-rich phases on demand -- with tunable kinetic correlations -- suggests the natural next steps of engineering monopole phases in finite-size arrays and in different lattice geometries. The additional availability of a wide-bandwidth dynamic probe opens the possibility of studying new regimes of magnetic charge dynamics in the more highly frustrated kagome systems, as well as the dynamics of the topological excitations recently demonstrated in vertex-frustrated ASIs~\cite{Nisoli2017}.  As a long-term prospect, the ability to field-tune the presence or absence of magnetic charges suggests the possibility of transistor-like devices based on monopole flow, realizing new potential applications for these emergent effective charges.

\section*{Acknowledgements} 
We gratefully acknowledge support from the Los Alamos LDRD program, help with SEM imaging from Chris Sheehan, and discussions with Clare Yu, Herve Carruzzo, and Gia-Wei Chern. Work at the NHMFL was supported by the National Science Foundation (NSF) DMR-1644779, the State of Florida, and the US Department of Energy. Work at Yale University was funded by the US Department of Energy, Office of Basic Energy Sciences, Materials Sciences and Engineering Division under Grant No. DE-SC0010778 and Grant No. DE-SC0020162.  Work at the University of Minnesota was supported by the NSF through DMR-1807124. 

\section*{APPENDIX: Methods}

\textbf{Sample fabrication.} ASI lattices were fabricated by methods similar to those employed in prior work~\cite{Zhang2013, Gilbert2016}. Briefly, electron beam lithography was used to pattern bilayer resist masks on Si/SiN substrates for subsequent metal deposition and lift-off. Islands of lateral dimension 220~nm $\times$ 80~nm were formed, with thickness $\sim$3.5~nm. Ultrahigh vacuum ($\sim$10$^{-10}$~Torr base pressure, $\sim$10$^{-9}$~Torr deposition pressure) electron beam evaporation at 0.05 nm/s was used for permalloy (Ni$_{0.8}$Fe$_{0.2}$) deposition, in a molecular beam epitaxy system. The islands were then capped with two layers of thermally-oxidized Al (total thickness $\sim$3~nm) to minimize oxidation of the permalloy.  

\textbf{Broadband magnetization noise spectroscopy.}  The ASI samples were mounted face-up in the $x-y$ plane, on a positioning stage that could be temperature-controlled from -10~$^\circ$C to +30~$^\circ$C. The horizontal and vertical islands were oriented along $\hat{x}$ and $\hat{y}$, respectively. The magneto-optical noise spectrometer is adapted from an instrument previously developed to measure out-of-plane magnetization fluctuations~\cite{Balk2018}. A weak probe laser ($<$1~mW), incident in the $x-z$ plane, was linearly polarized and focused to a small ($\simeq$4~$\mu$m diameter) spot on the ASI at $\simeq$45$^\circ$ incidence; $\sim$300 islands were therefore probed. Thermodynamic magnetization fluctuations along the $\hat{x}$ direction, $\delta m_x(t)$ (\textit{i.e.}, fluctuations of the horizontal islands only), imparted small Kerr rotation fluctuations $\delta \theta_K(t)$ on the polarization of the reflected laser, which were detected with balanced photodiodes. The magnetization noise was amplified, digitized, and its power spectrum was computed and signal-averaged in real time using fast Fourier transform (FFT) methods. Small coils were used to apply magnetic fields $B_x$ and $B_y$ in the sample plane.

The spectral density of the measured noise contained additional contributions from amplifier noise and photon shot noise, which were unrelated to and uncorrelated with the magnetization fluctuations from the ASI. We subtracted off these constant contributions by also measuring the noise spectra in the presence of large applied magnetic field ($B_x = B_y \simeq 20$~G) where all the islands were strongly polarized and magnetization noise from the ASI was entirely suppressed. 

To obtain the maps of the total (frequency-integrated) noise power \emph{vs.} applied magnetic field, for each value of magnetic field the noise spectrum was acquired for several seconds, which allowed us to record good quality data in the frequency range from a few hundred Hz to a few hundred kHz. For more detailed studies of the noise spectra over the broadest possible frequency range (shown for example in Fig. 5), the measured noise was signal-averaged for a longer time duration (typically tens of minutes), which increased the usable bandwidth from about 1~Hz to over 1~MHz.

\textbf{Monte Carlo (MC) spin dynamics simulations.} We performed standard Glauber MC simulations of conventional square ASI lattices with periodic boundary conditions. We used single spin updates (\textit{i.e.}, no cluster or loop moves), which should coarsely resemble the kinetics of the nanoislands in square ASI. Spins were chosen randomly, and the acceptance probability was $p=[1+\exp(\Delta/kT)]^{-1}$, where $\Delta$ is the usual energy difference associated with a spin flip and $k$ is the Boltzmann constant.  Typical simulations utilized $\sim$10$^6$ annealing steps, and then the magnetization was recorded for up to several millions of MC time steps at fixed temperature $T$ and applied field $B$. Noise spectra were computed directly from the time series via fast Fourier transform.

The simulations employed a vertex model where the energetics were obtained from two energy scales: the nearest-neighbor coupling $J_1$ between perpendicular spins, and the weaker next-nearest-neighbor coupling $J_2$ between collinear spins. The simulations used $J_1 / J_2 =1.8$, in line with previous micromagnetic simulations of ASI systems, for which the ratio varies from $1.4-2.0$ (depending on fabrication details).  Within this model, the energies of the four different vertex topologies in zero applied magnetic field were: $\epsilon_{\text{I}}=-4 J_1 +2 J_2$, $\epsilon_{\text{II}}=-2 J_2$, $\epsilon_{\text{III}}=0$, $\epsilon_{\text{IV}}=4 J_1 +2 J_2$. The vertex model defines clearly the applied in-plane magnetic fields at which the monopole plasma regime is realized -- namely, fields at which type-I and type-II vertices have equal energies: $|B_x|+|B_y| \propto \epsilon_{\text{II}}-\epsilon_{\text{I}}=4(J_1 - J_2 )$. Supplemental Material Fig. S1 shows the origin of the diamond-shaped noise maps and contains additional details.
 
The simulation temperature $T$ is defined in units of $J_2 /k$. Using $J_1 / J_2 =1.8$, the critical temperature $T_c$ below which square ASI spontaneously orders into its long-range type-I antiferromagnetic state is $T_c\simeq 2.4~J_2 / k$ (at zero applied magnetic field). To most closely match experimental conditions, MC simulations were typically performed at lower temperatures in the range of $T = 1.4 - 1.8~J_2/k$. Supplemental Material Fig. S4 shows additional simulations of the $T$- and $B$-dependent antiferromagnetic order parameter and specific heat in thermal square ASI, in relation to the ($T$-independent) monopole plasma regime. The applied fields $B_x$ and $B_y$ were defined in terms of the Zeeman energy on a single spin, and thus also in units of $J_2$. With these conventions, the monopole plasma was expected at $|B_x|+|B_y|=B_c = 3.2~J_2$.

\renewcommand{\thefigure}{S\arabic{figure}}
\setcounter{figure}{0}

\newpage

\begin{figure*}
\center
\section{Supplemental Material}
\includegraphics[width=1.2\columnwidth]{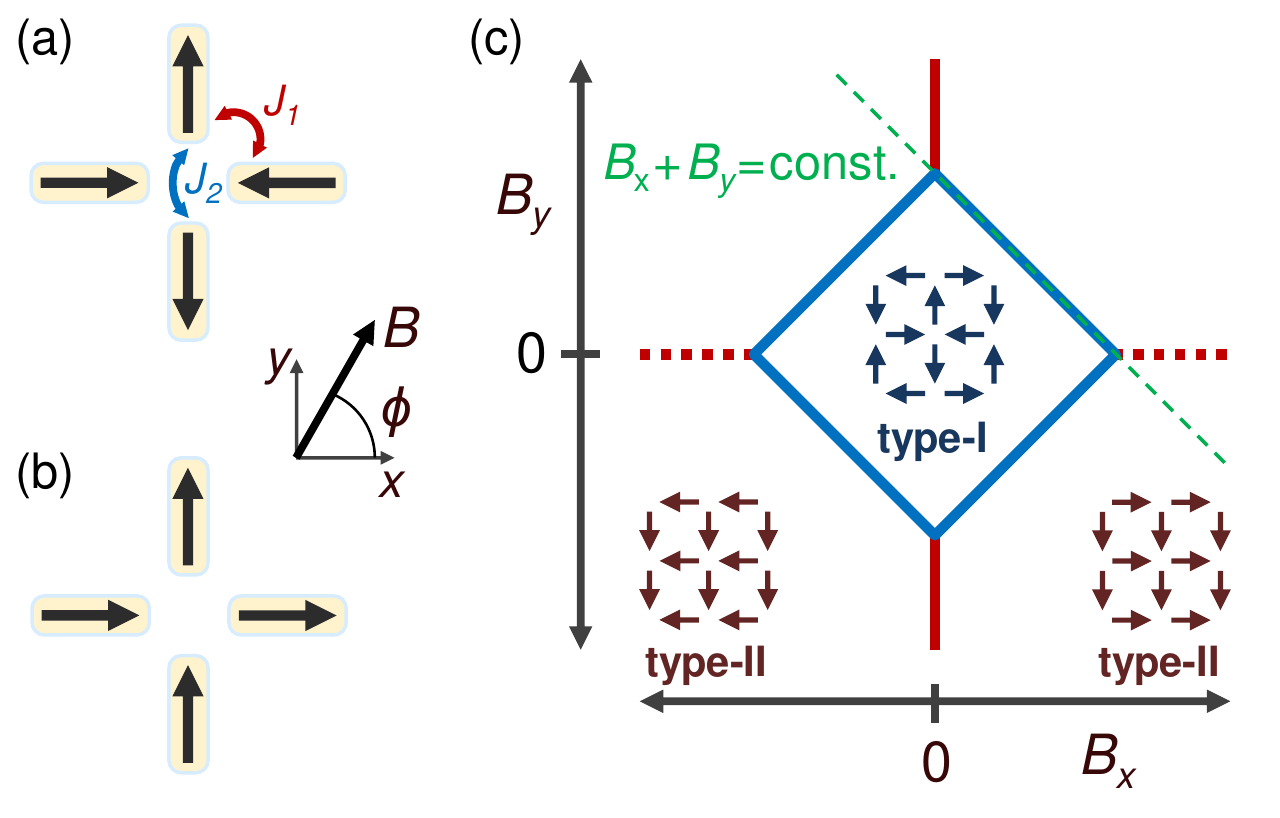}
\caption{Expected shape of the magnetic-field-dependent magnetization noise map in conventional thermal square ASI. (a), (b) Type-I and type-II vertices, respectively, in an in-plane magnetic field $\mathbf{B} = B_x \hat{x} + B_y \hat{y}$ applied at an angle $0 < \phi < 90^\circ$ with respect to the $x$ axis. The energies of the vertices ($\epsilon_{I}$ and $\epsilon_{II}$) are given by the nearest- and the next-nearest-neighbor coupling constants ($J_1$ and $J_2$, respectively) and the Zeeman energy due to $\mathbf{B}$, so that $\epsilon_I = -4J_1 + 2J_2$ and $\epsilon_{II} =  -2J_2 - \mu \left(B_x + B_y\right)$, where $\mu$ is the magnetic moment of a single nanoisland. (c) The expected shape of the magnetic field-dependent noise map, which is determined by the crossover magnetic field $B_c$ that is required to make type-I and type-II vertices energetically degenerate ($\epsilon_{I}=\epsilon_{II}$). For the type-II vertex shown in panel (b), $B_x + B_y =B_{c} = 4(J_1 - J_2)/\mu$, which is depicted by the dashed green line. Analogous expressions for the other three possible type-II vertices together define the diamond (blue line). Additionally, different orientations of type-II vertices are degenerate for $B_x = 0$ and large $B_y$ (or $B_y=0$ and large $B_x$) which results in the vertical (horizontal) ``tails'' of the diamond (where fluctuations along $\hat{x}$ ($\hat{y}$) are expected), depicted with solid (dashed) red lines.}
\label{diamond}
\end{figure*}

\newpage

\begin{figure*}
\center
\includegraphics[width=1.1\columnwidth]{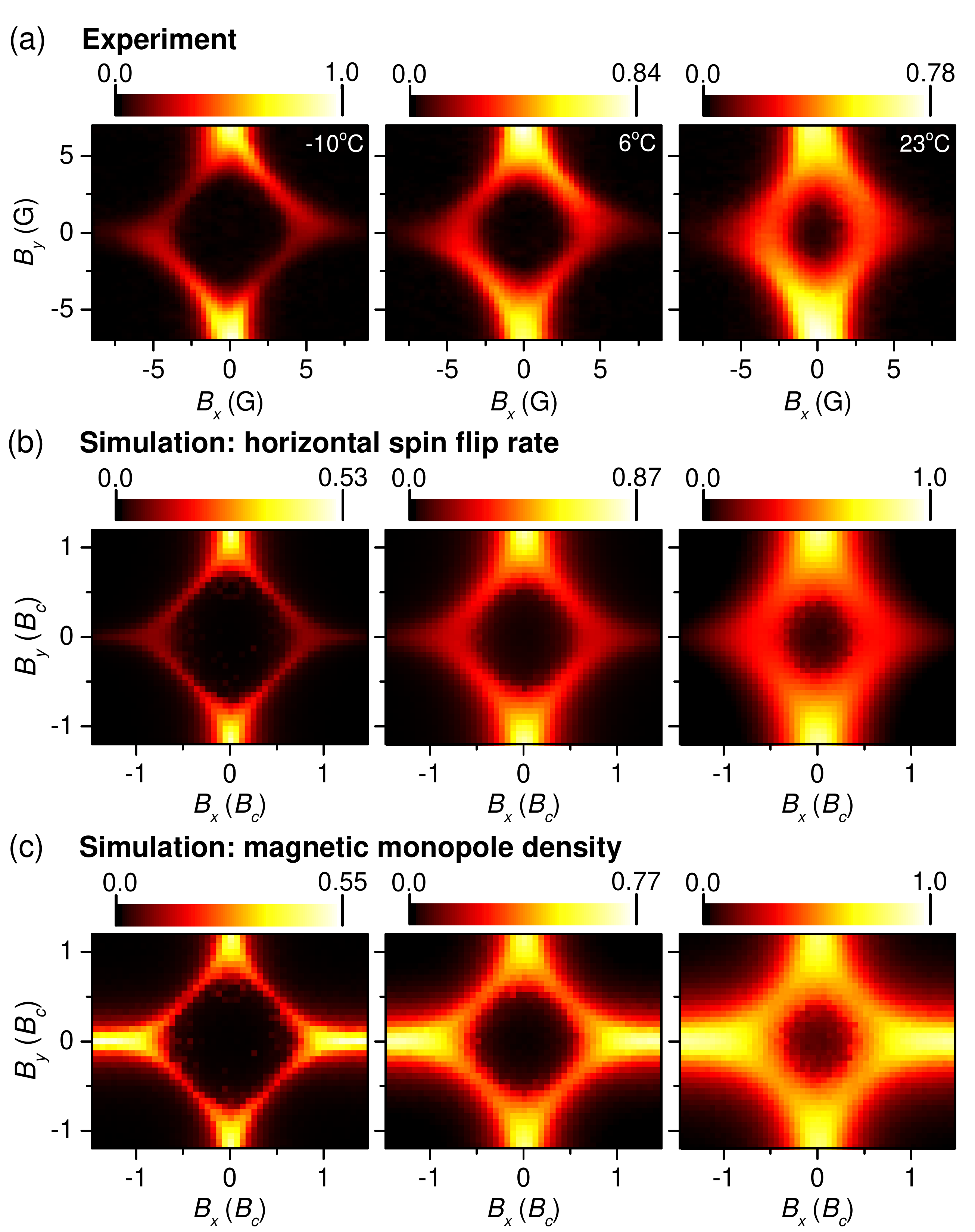}
\caption{Measured and simulated maps of the magnetization noise and magnetic monopole density at other temperatures. (a) Maps of the total (frequency integrated) magnetization noise from square ASI vs. $B_x$ and $B_y$, at --10$^{\circ}$C, 6$^{\circ}$C, and 23$^{\circ}$C. The boundary between the antiferromagnetic and polarized ordered states broadens at elevated temperatures where thermal fluctuations become more prominent. (b) Corresponding maps showing the calculated rate of horizontal magnetization flips from Monte Carlo simulations. (c) Maps of the calculated density of magnetic monopoles (type-III vertices) from simulations. Regions of high monopole density are also the most active. The four-fold symmetry of the monopole density map is not captured by the experiment, which is sensitive only to fluctuations along $\hat{x}$ but not $\hat{y}$.}
\label{maps_vs_temp}
\end{figure*}

\newpage

\begin{figure*}
\center
\includegraphics[width=.9\columnwidth]{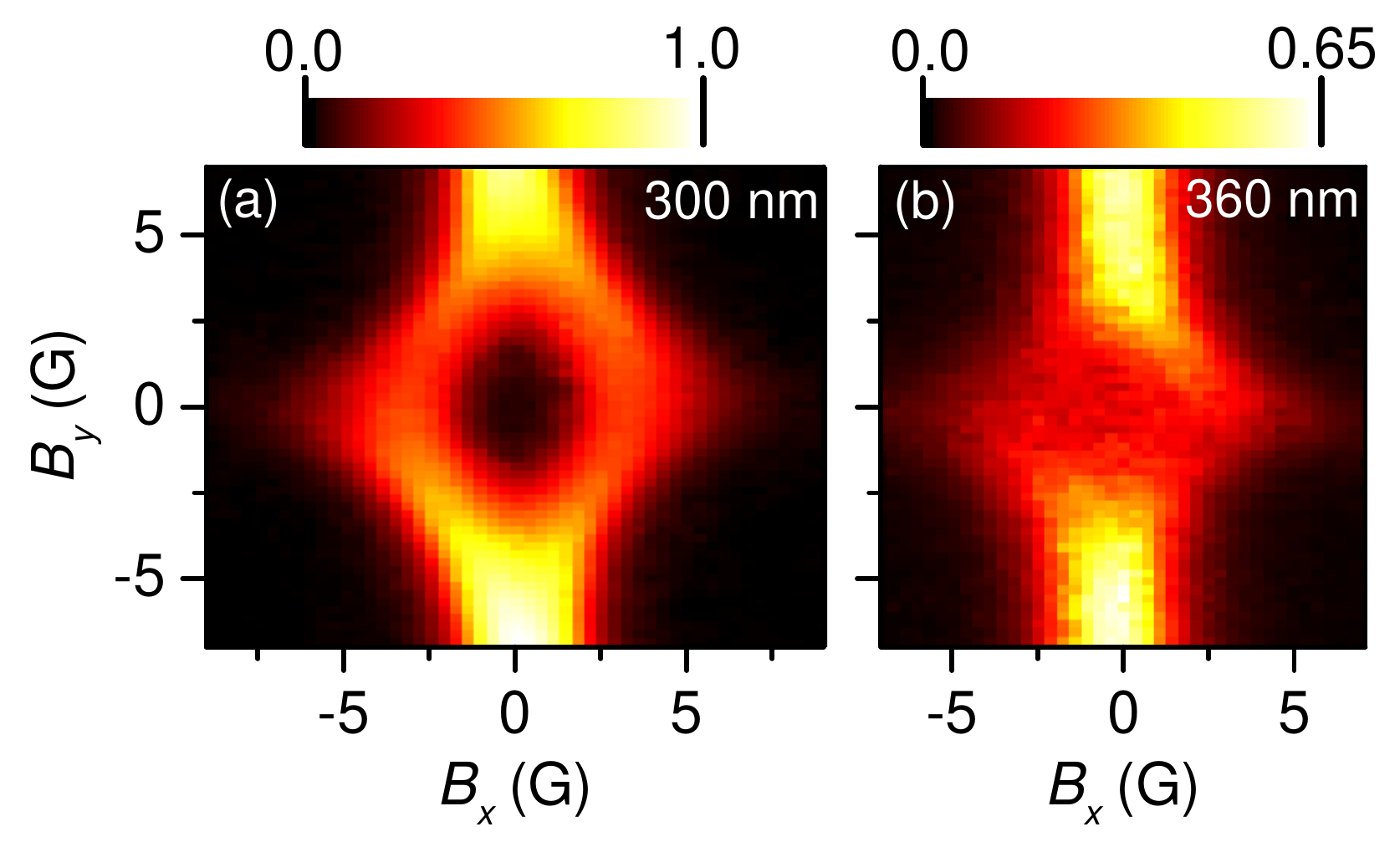}
\caption{Measured maps of the magnetization noise for square thermal ASI with different lattice constants. (a) Map of the total (frequency integrated) magnetization noise measured at $23^\circ$C for the square ASI made of 220~nm $\times$ 80~nm islands, with the lattice constant $d = 300$~nm (\textit{i.e.}, the same ASI used for all the experiments described in the main text). (b) Analogous map for square ASI with $d=360$~nm. As expected, the diamond-shaped boundary is still visible, but appears at smaller values of the applied field, due to weaker coupling between the islands.}
\label{maps_vs_lattice}
\end{figure*}

\newpage

\begin{figure*}
\center
\includegraphics[width=1.7\columnwidth]{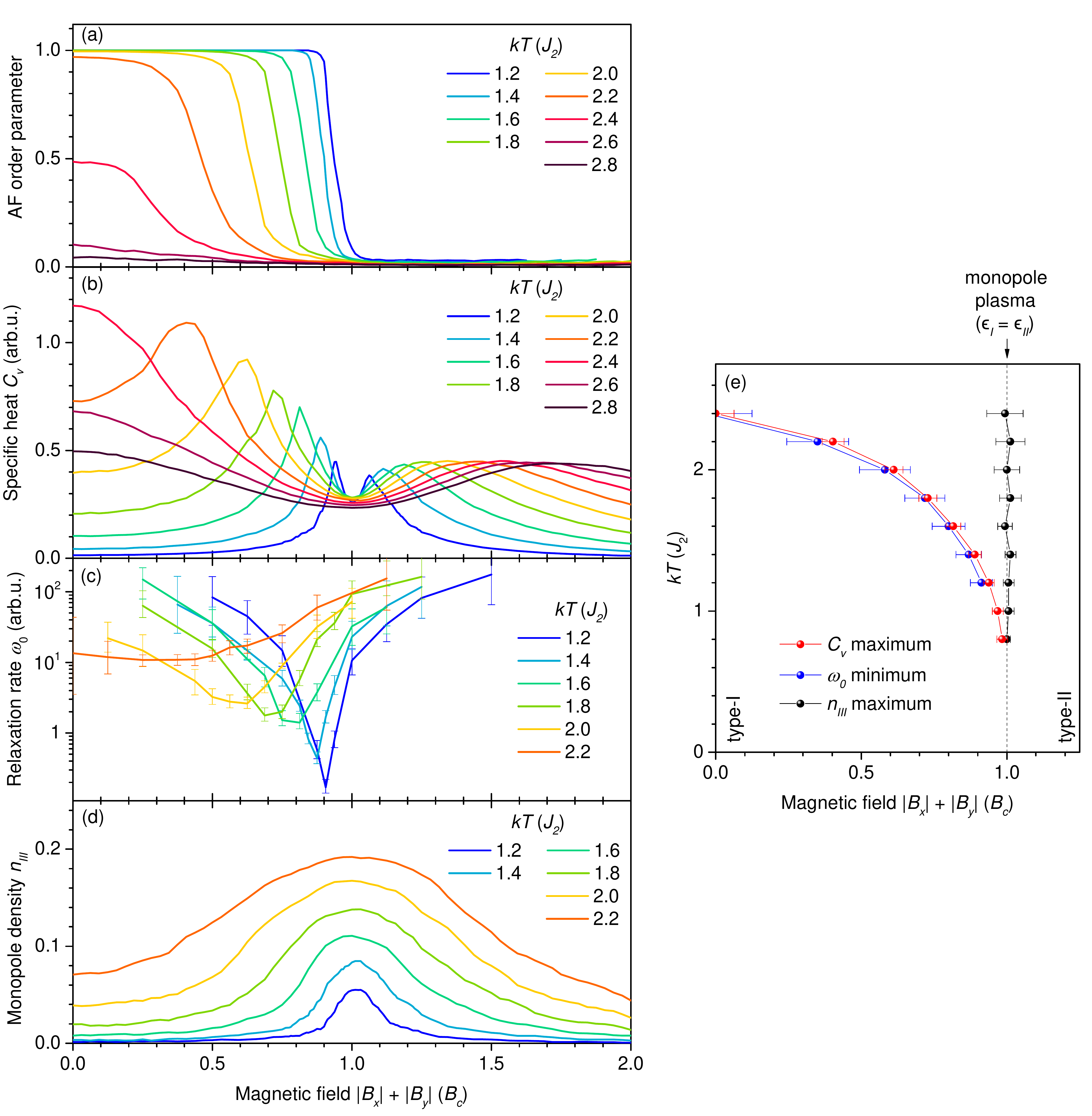}
\caption{Monte Carlo simulations of thermal square ASI: visualizing the monopole plasma regime in relation to the antiferromagnetic phase transition. (a) Dependence of the  antiferromagnetic (AF) order parameter on in-plane field $B$ applied along a lattice diagonal ($B_x = B_y$), at different temperatures $T$.  At $B$=0, square ASI exhibits long-range AF (type-I) order below a critical temperature $T_c \approx 2.4~J_2 / k$ (\textit{i.e.}, the system orders spontaneously and is dominated by one large AF domain). When $B>0$, the AF transition occurs at lower temperatures. Equivalently, AF order is suppressed when $B > B_{\mathrm{AF}}$, where $B_{\mathrm{AF}}$ is the temperature-dependent transition field (see also the phase boundary in panel (e)). At $T=0$, $B_{\mathrm{AF}} = B_c$, the crossover field at which type-I and type-II vertices have equal energy. (b) Calculated specific heat $C_v$ of square ASI vs. $B$, at different temperatures. The first maximum of $C_v$ corresponds to the AF phase transition at $B_{\mathrm{AF}}$. $C_v$ exhibits a minimum at the temperature-independent crossover field $B_c$, where the monopole plasma exists. The second maximum of $C_v$ at $B>B_c$ is due to a Schottky-type anomaly arising from the increasing energy difference between type-I and -II vertices. (c) Characteristic relaxation rates $\omega_0$ extracted from simulated noise spectra. The minimum of $\omega_0$ is related to the well-known slowing-down of kinetics at phase transitions, and occurs at $B_{\mathrm{AF}}$. (d) Calculated density of type-III monopole vertices, $n_{III}$. The maximum reveals the monopole plasma regime, which always occurs at $B_c$, independent of temperature. (e) The $T$- and $B$-dependent AF phase boundary of square ASI, plotted along with with the maximum monopole density.}
\label{order}
\end{figure*}

\end{document}